\RequirePackage{fix-cm}
\documentclass[smallextended]{svjour3}       
\smartqed  
\usepackage{graphicx}
\usepackage[round]{natbib}

\journalname{Human Ecology}
\begin{document}

\title{Non-L\'evy ﻿mobility patterns of Mexican Me'Phaa peasants searching for fuelwood
}


\author{Octavio Miramontes \and Og DeSouza \and Diego Hern\'andez \and Eliane Ceccon}


\institute{O. Miramontes \at
              Instituto de F\'isica and\\ 
              Centro de Ciencias de la Complejidad,\\ 
              Universidad Nacional Aut\'onoma de M\'exico,\\
              Ciudad Universitaria 04510 DF, MEXICO\\ 
              Tel.: +52-55-56225014\\
              \email{octavio@fisica.unam.mx}             
              \\ \emph{Present address of O. Miramontes:} Departamento de F\'isica, UFPR, Curitiba 81530-900, BRAZIL.
           \and
           O. DeSouza \at
           Laborat\'orio de Termitologia, Entomologia,\\
           Universidade Federal de Vi\c{c}osa, \\
           Vi\c{c}osa, MG, BRAZIL\\
           \and
           D. Hern\'andez \at
           Posgrado en Ciencias Biol\'ogicas,\\
           Facultad de Ciencias, \\
           Universidad Nacional Aut\'onoma de M\'exico,\\
           Ciudad Universitaria 04510 DF, MEXICO\\  
	   \and
           E. Ceccon  \at
              Centro Regional de Investigaciones Multidisciplinarias,\\
              Universidad Nacional Aut\'onoma de M\'exico,\\
              Av. Universidad s/n, Circuito 2,
              62210, Col. Chamilpa,\\
              Cuernavaca, Morelos, MEXICO        
\\ \emph{Present address of E. Ceccon:} Departamento de Biolog\'ia, UFPR, Curitiba 81530-900, BRAZIL.
}

\date{Received: date / Accepted: date}

\maketitle

\begin{abstract}
We measured mobility patterns that describe walking trajectories of individual Me'Phaa  peasants searching and collecting fuelwood in the forests of ``La Monta\~na de Guerrero" in Mexico. These one-day excursions typically follow a mixed pattern of nearly-constant steps when individuals displace from their homes towards potential collecting sites and a mixed pattern of steps of different lengths when actually searching for fallen wood in the forest. Displacements in the searching phase seem not to be compatible with L\'evy flights described by power-laws with optimal scaling exponents. These findings however can be interpreted in the light of deterministic searching on heavily degraded landscapes where the interaction of the individuals with their scarce environment produces alternative searching strategies than the expected L\'evy flights. These results have important implications for future management and restoration of degraded forests and the improvement of the ecological services they may provide to their inhabitants.

\end{abstract}

\section*{Introduction}
\label{intro}

Mobility patterns of humans searching for resources seem to follow specific statistics known to physicists as anomalous diffusion \citep{klafler2005anomalous}. From the prehistoric tribes in Europe searching for quality stone quarrels \citep{brantingham2006measuring} to the hunter-gatherers of Africa \citep{brown2007levy, brown2010hunter}, it is becoming clear that distances traversed when humans travel, migrate, disperse and explore a territory is not quite a Gaussian stochastic process but one described by scale-free statistics \citep{brockmann2006scaling,flores2007dispersal, gonzalez2008understanding, rhee2011levy}.

Understanding the non-Gaussian mobility nature of humans is very important for realistically explaining diverse patterns in humans such as mating (gene flow), cultural changes, foraging, spread of disease and migration processes, all of which have potential and unexpected impacts on the daily life of individuals and the complexity of societies. Mobility drives the dynamics of encounters among humans and influences the way vital resources and ecosystem services are used, for example food and energy sources. In this study, we measured the patterns of displacements of individual peasants when searching and collecting fuelwood in the forests of ``La Monta\~na de Guerrero" in the southern Mexican state of Guerrero. We discuss evidence that these patterns do not conform to L\'evy flight statistics dominated by power-law probability distributions of the form $P(l)\sim l^{-\mu} $ with scaling exponent $\mu$ ($1<\mu\leq3$) and $l$ being the length of the displacements. The absence of L\'evy flight patterns is not, however, evidence against L\'evy foraging as a human mobility phenomenon but rather that, in this case, it is the result of foragers interacting with an environment characterized by the scarcity typical of degraded landscapes.   

Conventional optimal foraging theory states that animals and humans search for resources guided by decisions made in order to maximize a given quantity, usually the net energy gain \citep{charnov1976optimal, stephens1986foraging}. The efficiency of the search is then in proportion to the behavioral repertoire of the foragers who may employ, to varying degrees, cognitive tools such as memory and information processing when facing the searching space. Most searching models fall in two main groups: (1) random searching where the forager does not have previous information of the prey field such as the spatial distribution or density and (2) deterministic searching where the forager is able to use memory maps of the territory and therefore has relatively good previous information about the prey field richness and location.  

\subsection*{Random searching}

Random searching is by no means exclusively performed by simple low-profile organisms \citep{matthaus2011origin, hays2011high}, it may be used even by humans under specific environment constraints, for example when searching for a plane that has crashed in the open ocean at an unknown location. It is an optimal strategy when there is no previous information about where the target or prey is located. Recently, optimal random searching theory has been developed further around the so called L\'evy flights \citep{shlesinger1986lévy, chechkin2008intro, viswanathan2011physics}. It has been demonstrated that step-lengths generated by a power-law probability distribution is an optimal searching strategy when the scaling exponent is $\mu\approx1$ when the prey is removed (destructive scenario) or $\mu=2$ when the prey is not removed (non-destructive scenario) \citep{viswanathan1999optimizing}. 

In the destructive scenario there is no reason to re-visit previous sites such that this searching strategy is conveniently done by ``ballistic motion" where successive step-lengths are very large in size and are only separated by a random change of angle. In practice, it corresponds to the strategy of ``follow a straight line until you find something". Ballistic searching was first suggested to exist in the very long rectilinear flights of the wandering albatross but was later called into question \citep[but see also \cite{boyer2008evidence}]{edwards2007revisiting}. A non-destructive searching scenario, in which the prey can regenerate or reappear in the same location at a later time, is optimally performed when the forager alternates a few long steps with more abundant short steps. This scenario with $\mu\sim2$ was found to exist in many  biological examples but has also been recently reviewed and questioned citing methodological inadequacies \citep{james2011assessing}. Since recent unquestioned examples of L\'evy flights are continuously being published \citep{sims2008scaling, hays2011high}, the state of the art of this research area has attracted much interest and is rapidly changing.     

\subsection*{Deterministic searching}

Animals and particularly humans may rely on cognitive tools and memory maps in order to perform more efficient searches on a territory previously known or similar to one previously known. A common strategy is ``go to the nearest richest place" in which the forager makes a decision on where to go by weighing the profit offered by the site against the energy invested in reaching it, depending on the forager's present location. A well-studied problem related to this deterministic searching mode is the tourist problem in which a tourist wishes to visit a number of cities without repeating them and where some cities are more attractive than others, posing a challenging optimization problem \citep{stanley2001salesman, lima2001deterministic,campiteli2006image}. Scale-free movement statistics are known to be an emergent property in models of decision-making intelligent agents performing deterministic searches in heterogeneous environments \citep{boyer2005looking, boyer2009levy, boyer2010modelling}. Emergent scale-free mobility patterns have been suggested to exist in animals ranging from microzooplankton to spider monkeys, among others \citep{reynolds2008deterministic, boyer2004modeling, boyer2006scale}. 

In the case of the spider monkeys search patterns \citep{ramos2004lévy} it has been suggested that these arise as a consequence of the interaction of the searching individuals with the environment in such a way that characteristics like density and spatial richness distribution influence the searching strategy \citep{boyer2006scale}. Different searching options may be displayed in different quality habitats. In rich habitats, resources such as food are so abundant at close ranges that long-distance displacements are not expected. At the other extreme, when the environment is poor with scarce resources, there is no point in traveling far away because the energy invested in long trips is not justified since the gain will be the same that when moving locally. Again long-distance displacements are not expected. The situation is very different when the environment is neither exceptionally rich nor poor. In this case L\'evy like displacement statistics are robustly expected to emerge \citep{boyer2006scale, boyer2009levy}.

Levy-like statistics have unexpected consequences for social foragers. For example, in spider monkeys, it has been shown that complex social structures spontaneously emerge from the interactions of individuals moving under such conditions. These social networking even have a small-world dynamics \citep{ramos2006complex}. In the case of the Dobe Ju/’hoansi hunter-gatherers mobility patterns, it was suggested that the L\'evy foraging influenced the way tribes visited and remained for a time in specific foraging sites. When individuals from different tribes came into contact in these sites, cultural exchanges and inter-tribe couple mating (gene flow) took place. In this way a complex social structure emerged paired to the local use of resources \citep{brown2007levy}. 

\section*{The Me'Phaa of La Monta\~na de Guerrero}

Me'Phaa is a pre-Columbian indigenous ethnic group that inhabits the region known as ``La Monta\~na de Guerrero" (MG) in the state of Guerrero in Southwestern Mexico. Me'Phaa towns scatter the area around the city of Ayutla de los Libres (16$^\circ$54' N, 99$^\circ$13' W). The Me'Phaa were until very recently knew as ``Tlapanecas'' which is pejorative Nahuatl term from the time the Aztec Empire ruled the region. MG is one of the poorest and least developed areas in Mexico, with Human Development Index (HDI) of around 3.2 \citep{morales2006remesas, taniguchi2011guerrero}, comparable to some areas of sub-Saharan Africa despite its close proximity to the world-class resort of Acapulco. A typical Me'Phaa individual has no access to health services, schools, paved roads, telecommunications, or electricity, and their situation has been worsened by the recent incidence of militarization, social conflict, and violence \citep{hebert2006ni, camacho2007montana}. 

\subsection*{Fuelwood usage and availability}

\begin{figure*}
  \includegraphics[width=1.0\textwidth]{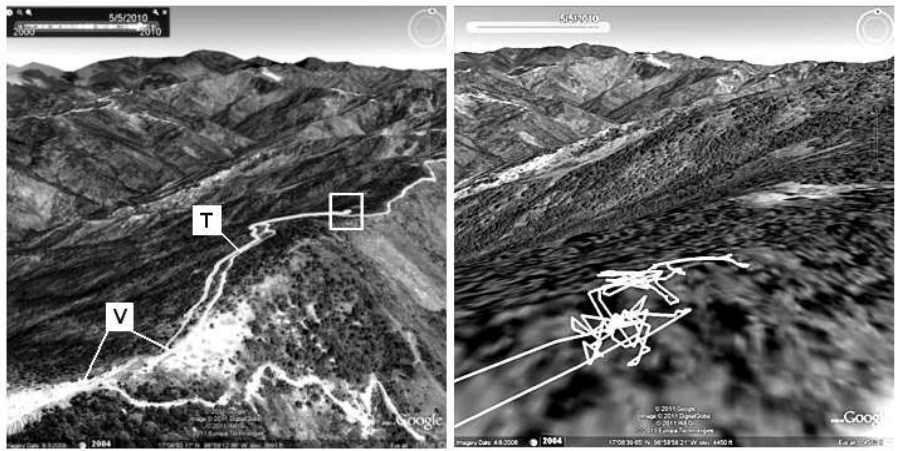}
\caption{Panoramic view of the landscape of ``La Monta\~na de Guerrero" in Mexico, one of the poorest regions in the country. The image at the left shows the hilly nature of the seasonally tropical dry forest currently characterized by large rates of deforestation and soil erosion. The image shows one example of a walking trajectory (T) performed by a GPS-equipped peasant when traveling to collect fuelwood. The area of large white spots at the bottom of the image is a village (V) of Me'Phaa peasants of about 20 houses (17.1388969 N, -98.9851532 W,  1314 msl). At the end of the recorded trajectory there is a shift in the searching behavior that becomes an area-restricted active search. The behavior is shown in the square window that has been enlarged in the image at the right. This last image shows in detail, the actual displacements over the terrain when searching. Images are an overlap of the GPS positions on Google Earth\texttrademark imagery. Images courtesy of Google Inc.}
\label{fig:1}       
\end{figure*}

The landscape of MG is hilly with forest patches mostly on hilltops (see Figure~\ref{fig:1}). Seasonally tropical dry forest is the most prevalent type of vegetation, but oak and pine forests are present in the highlands (around 1890 msl). The environment is severely degraded with high erosion and deforestation rates \citep{landa1997}. MG is known to have been subject to occupation and anthropogenic impact for as long as 2000 years \citep{berrio2006environmental}, which has caused the present pattern of degradation. Fuelwood is mostly composed of species belonging to the genera \emph{Acacia}, \emph{Leucaena}, \emph{Lysiloma}, \emph{Prosopis} and \emph{Pithecellobium} \citep{cervantes1998growth}. \emph{Quercus} and \emph{Pinus} species are also collected when the searching is done in the highlands. 

Many Me'Phaa peasants do not have any access to modern energy sources and most continue to rely primarily on fuel wood for low-tech residential uses such as cooking and heating. MG is regarded as one of the main hot spots of fuelwood usage in Mexico and so requires priority attention since this practice continues to stress the already degraded forest \citep{Ghilardi2007475}.  

\section*{Material and methods}
\label{sec:3}

Twelve voluntary Me'Phaa peasants were trained to operate GPS data loggers in order to register positions along a trajectory in the field when searching for fallen fuelwood. The learning process posed no difficulty and after few minutes of manipulating the devices the volunteers showed adequate skill when operating them. Specific instructions were given to switch on the devices when leaving home on a search and collection excursion and to switch them off when returning home. The devices were configured to log data automatically as soon as the first accurate position fix was obtained.  The GPS data loggers were worn on a neck lanyard and protected by a plastic bag from dust and rain. All volunteers were males in their 20s to 50s and the searching for fuelwood was done alone with no supervision or accompaniment by the research team. The peasants were left on their own to carry out the searches whenever and wherever they do so normally. Only fallen wood was collected, and no trees were felled. Normally, fallen wood is picked up and carried by hand until a sufficient amount is collected and tied together into a bundle carried over the shoulders. All volunteers belonged to NGO for organic agriculture production based in the city of Ayutla de los Libres (Xuajin Me´Phaa AC). Participants live in scattered communities in the mountains surrounding the city, separated from each other by several kilometers. 

We used Holux-M241 data loggers (Holux Technology Inc.) equipped with a MKT GPS-chipset capable of storing 100,000 points. The devices were set to record successive points at intervals of 10 seconds. All devices were recovered and the data transferred to a computer for analysis. The data was analyzed with an \emph{ad-hoc} program we developed that implemented a Haversine algorithm to calculate distances between two geo-referenced points on earth. The participants recorded a total of 114 field trips, but only 10 of them contained search-related data. These recordings contained a total aggregated amount of 3386 displacements, from which 1231 displacements corresponded to searching displacements  with a maximum value of 28.4 meters. All displacements less than 1 meter were ignored when analyzing the displacements since such distance in 10 seconds is approximate to a waiting interval. These values however were considered when analyzing the waiting times.

All statistical analysis was conducted using the R Open Source Statistical Language \citep{rmanual} on a Linux machine (Ubuntu 11.04). Step-lengths and waiting-times distributions were fitted with two alternative statistical models \citep{newman2005power,edwards2007revisiting}: a power-law $l^{-\mu}$ and the exponential $e^{-\lambda l}$. Maximum Likelihood Estimation (MLE) was used for evaluating model parameters and loglikehood values.  In a set ${\bf x}$ of random variables of size $n$: ${\bf x}=\{x_1,x_2,\dots,x_n\}$, the scaling exponent $\mu$ of the power-law is giving by:

\begin{equation}
\mu = 1 + n \left[\sum_{i=1}^{n} ln\frac{x_i}{x_{a}}\right]^{-1}
\end{equation}

In the case of the exponential distribution, the $\lambda$ parameter value is given by:

\begin{equation}
\lambda = \left[\sum_{i=1}^{n} \frac{x_i}{n-x_{a}}\right]^{-1}
\end{equation}

In both cases $x_a$ is the minimal value of ${\bf x}$ from which the probability distribution hold; in our case $x_a$ was fixed as the minimun $x$ value, so that the entire time-series was evaluated. A Model Selection approach involving an Akaike Information Criterion (AIC) was conducted in order to identify among the two statistical models the one that better explain the data \citep{burnham2002model}. The AIC of model $i$ is defined as $AIC_{i} =-2L_i+2D_i$, where $L$ is the log-likehood of the fit and $D$ is the number of parameters of the model. The model with minimal value of AIC is regarded as the most parsimonious and is normally considered the most likely to explain the data \citep{burnham2002model}.

\section*{Results}
\label{sec:4}

\begin{figure}
  \includegraphics[width=0.5\textwidth]{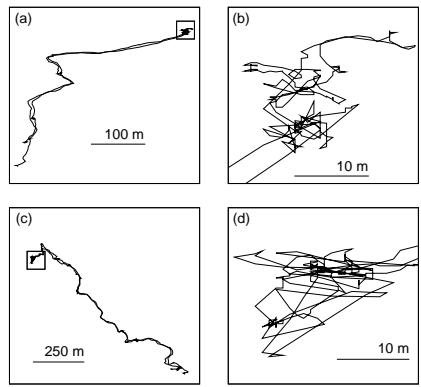}
\caption{Two examples of travel and search trajectories as recorded every ten seconds by walking Me'Phaa peasants (a and c). These images show in detail the two phases of the searching and collecting of fuelwood. The first phase is a simple ``ballistic" displacement towards the potential collecting site and may span up to several kilometers. When the  peasants arrive to the site of interest, the behavior is shifted as shown in the two enlarged figures (b and d). The pattern of mobility is then replaced by an entangled succession of many short steps and few large steps.}
\label{fig:2}       
\end{figure}

The one-day search and collecting excursions typically follow a mixed pattern of nearly-constant steps when individuals displace from their homes towards the forest (a ``ballistic phase") and a pattern of steps with a fat-tail like distribution when actually doing the searching for fallen wood (see Figure~\ref{fig:2}). This pattern of mobility behavior is very common in nature and has received several names \citep{Knoppien1985273, lomholt2008lévy, bénichou2011intermittent}. It is characterized by an adaptive switching between searching behaviors depending on the availability and location of the targets. The first phase is dominated by a simple displacement of the forager to the area of interest and a second phase follows when a potentially resource-rich location is reached. In this 
second phase, active and careful search behavior is initiated.  We ignored the displacements pertaining to the first ``ballistic" transportation phase and analyzed only the displacements in the second active search phase. 

\begin{figure}
  \includegraphics[width=0.5\textwidth]{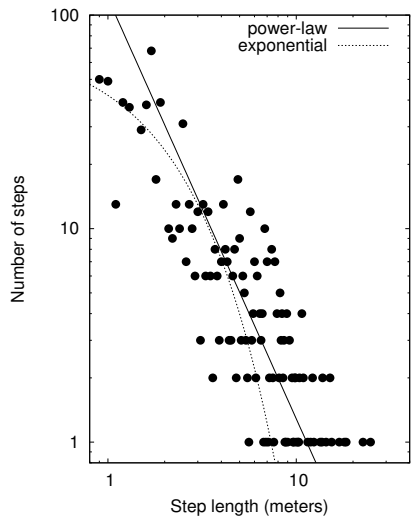}
\caption{Statistics of the step lengths. This is a log-log plot of the histogram of the distances traveled by the peasants versus the number of such distances, recorded every ten seconds. The straight line shows the power-law fit as estimated using the MLE method. The power-law has an scaling exponent $\mu=1.97$. The curved line is the exponential tested as an alternative model over the same interval (see Table~\ref{tab:1}). The exponential model better explains the data as suggested by an AIC model selection.}
\label{fig:3}       
\end{figure}

\begin{figure}
  \includegraphics[width=0.5\textwidth]{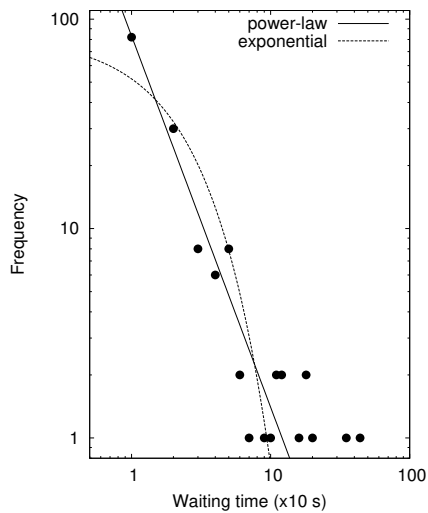}
\caption{Statistics of waiting times. This is the log-log plot of the histogram of the waiting times versus their frequency, recorded every ten seconds. The straight line shows the power-law fit as estimated using the MLE method. The power-law has an scaling exponent $\mu=2.66$. The curved line is the exponential tested as an alternative model over the same interval (see Table~\ref{tab:1}). The AIC model selection suggests that the power-law is the most adequate model to explain the data.}
\label{fig:4}       
\end{figure}

Step-lengths distribution was fitted with two alternative statistical models using Maximum Likelihood Estimation, a power-law and the exponential. The parameter values found are giving in Table~\ref{tab:1} (see Figure~\ref{fig:3}). A Model Selection approach points to the exponential model as the most parsimonious, having the lower AIC value. On this basis, we conclude that the Me'Phaa searching displacements are essentially a Brownian-like motion. On the other hand, the waiting times, that is the time intervals with no displacements, were analyzed with the same procedure as above (see Figure~\ref{fig:4} and Table~\ref{tab:1}). In this case the lower AIC value suggests that the most likely model explaining the data is the power-law model ($\mu=2.6$), meaning that the resource is scattered in patches with amounts of fuelwood following a scale-free probability distribution, a result that is in agreement with the known fact that the size distribution of forest trees and their associated biomass is scale-invariant \citep{enquist2001invariant}.

\section*{Discussion and summary}
\label{sec:5}

Mobility is increasingly becoming a problem of interest for anthropology \citep{brown2007levy, richerson2008being}. Of particular interest is how individual foraging movements influence the pattern of social contacts that in turn determine the emergence of complex social structures, in humans \citep{brown2007levy} and non-human primates \citep{ramos2006complex}. The patterns of social contacts are also important for understanding the spatial dynamics of social phenomena such as cultural and social changes \citep{richerson2008being}, gene flow \citep{slatkin1973gene,sokal1989spatial, wakeley1999nonequilibrium}, disease spreading and vaccination strategies \citep{miramontes2002dynamical, mao2010dynamic}, among others. Human mobility and the environment are also entangled in a complex matrix of feedback interactions where mobility directly impacts the availability of resources and specific ecosystem services at various scales and where environmental fluctuations and landscape degradation may also negatively influence the patterns of human mobility and social behavior, even enhancing social conflicts \citep{homer1994environmental, raleigh2007climate, burke2009warming, Hsiang2011civil}.

For many years, it was thought that human migrations and individual mobility could be described by Gaussian probability distributions; however recent studies suggest that human mobility may be better explained by anomalous diffusion where the statistics of displacements follow power-law distributions in the form of L\'evy flights. The origins of L\'evy statistics in human displacement is therefore an issue of 
increasing interest. Due to the fact that humans search using cognitive tools for decision-making, most of their foraging behavior is aimed at optimizing a cost/benefit ratio, as conventional foraging theory predicts \citep{charnov1976optimal}. Such optimization is at the core of a modern approach that argues for the existence of a deterministic behavior of searching where the spatial distributions of the target field (richness and density) cause the spontaneous emergence of the L\'evy foraging in intelligent agents \citep{santos2007origin, boyer2008intricate, boyer2009levy}. When the environment is scarce in resources, deterministic searching force non-L\'evy patterns of displacements because the L\'evy stable distributions start converging into a Gaussian one when the scaling exponent $\mu$ of the power law is $> 3$ \citep{chechkin2008intro, Nurzaman2011from}. Such non-L\'evy patterns are the best local strategy for such environments but these are non-global optimal solutions.

It is estimated that nearly 2.5 billion people in developing countries world-wide make use of fuelwood in order to meet their residential energy needs \citep{international2009world}. Most of these people live in areas subject to strong environmental pressure. This is the case of the Mexican Me'Phaa peasants in ``La Monta\~na de Guerrero" in Mexico. This impoverished indigenous group has inhabited the area since pre-Columbian times and so have made use of the ecosystem services since then with non-persistent large-scale management strategies. We have studied the mobility patterns of the Me'Phaa peasants when searching and collecting fuelwood in the field. We have found a mixture of foraging behaviors that consist of a long trip (up to several kilometers on foot) from their homes to the collecting sites. This phase is composed of nearly constant distance steps that mostly follow the path of roads or paths. This is similar to a ``ballistic behavior" where there is no active searching at all. When arriving to an interesting site that contains a significant amount of fuelwood, the movement behavior is replaced by an active search composed of a mixture of abundant small steps alternated with few long steps in a fashion reminiscent of intermittent searching \citep{Knoppien1985273, lomholt2008lévy, bénichou2011intermittent}. The statistical distribution of the step lengths is explained better by an exponential model. 

What is the meaning and the origin of an apparently non-optimal Brownian-like movement pattern in the Me'Phaa searching process? First we should examine in more detail the searching behavior as performed by the individuals. When arriving in the area to be searched, peasants do have a fairly good view of where the fallen wood lies around since the density of trees is low in a degraded seasonally tropical dry forest (also, there are no other major physical obstacles). This means that the individuals would not move around searching randomly as if blind. Instead, they will move to where they see (tens of meters) there is fallen wood of good size and in good amount. This searching behavior is then repeated and it matches the pattern of a deterministic search as explained in the introduction. We rule out the behavior leading to a L\'evy distribution with optimal scaling exponents $\mu=1$ or $\mu=2$, typical of random searching scenarios \citep{viswanathan2011physics}. Instead we emphasize that, in models of deterministic searching, it has been argued already that environments with scarcity of resources may lead to Brownian-like displacements. This does not mean that the search is inefficient but that a Brownian-like searching pattern becomes locally optimal under such extreme conditions \citep{boyer2006scale, boyer2009levy}.

Brownian-like behavior is also known to occur in other non-human primates \citep{Schreier201075, sueur2011non}, stressing the importance of considering the role of the environment in influencing the mobility patterns of foragers. Spatial distribution, density and resource abundance can not be separated from the study of forager mobility \citep{de2011levy, miramontes2011fractal}, especially when these involve decision-making and optimization efforts \citep{boyer2009levy}. 

Fuelwood search and collection by the Mexican Me'Phaa peasants is characterized by large amounts of energy invested in traversing large distances walking in order to profit poorly, since the collection is a low-tech activity limited by the carrying capacity of the individuals. Therefore the searching behavior of the Me'Phaa despite of involving elaborate decision-making and complex interactions with the environment is far from a global optimal solution. This situation may be reversed or improved if Me'Phaa peasants could actively modify the scarce nature of their forests by means of adopting forest restoration practices that would increase the availability of wood close to their homes. Research on the restoration of ecosystem services in the region is an activity in progress following this study.

%
%

\begin{table}
\caption{MLE parameter values of two models}
\label{tab:1}       
\begin{tabular}{llll}
\hline\noalign{\smallskip}
Model & Parameter & Loglike & AIC  \\
\hline\noalign{\smallskip}
\emph{Step-lengths} & & & \\
\noalign{\smallskip}\hline\noalign{\smallskip}
Power-law & $\mu=1.97$ & $-3336.00$ & $6674.00$ \\
Exponential & $\lambda=0.03$ &$-3274.37$ & ${\bf 6550.75}$\\
\noalign{\smallskip}\hline\noalign{\smallskip}
\emph{Waiting-times} & & &\\
\noalign{\smallskip}\hline\noalign{\smallskip}
Power-law & $\mu=2.66$ & $-162.44$ & ${\bf 326.89}$\\
Exponential & $\lambda=0.47$ &$-261.01$ & $524.03$\\
\noalign{\smallskip}\hline
\end{tabular}
\end{table}

\begin{acknowledgements}
We very much appreciate PAPIIT-UNAM Grants IN-118306, IN-107309 and IN-304409, PASPA-DGAPA grants, a CONACYT-CNPq Bi-national Joint Project on the Dynamics of Mexico-Brazil Tropical Forests and the Centro de Ciencias de la Complejidad (C3) for financial support. ODS is supported by a fellowship from Brazilian National Council for Research (CNPq 302486/2010-0). OM and EC thanks the Universidade Federal de Paran\'a in Brazil for hosting a sabbatical leave. We thank Pedro Miramontes and Lynna Kiere for useful comments. Special thanks to all the Me'Phaa volunters and the Xuajin Me´Phaa AC ONG.
\end{acknowledgements}

\bibliographystyle{spbasic}      

\begin{thebibliography}{1}

\bibitem[{B{\'e}nichou et~al(2011)B{\'e}nichou, Loverdo, Moreau, and
  Voituriez}]{bénichou2011intermittent}
B{\'e}nichou O, Loverdo C, Moreau M, Voituriez R (2011) Intermittent search
  strategies. Reviews of Modern Physics 83(1):81

\bibitem[{Berr{\'\i}o et~al(2006)Berr{\'\i}o, Hooghiemstra, van Geel, and
  Ludlow-Wiechers}]{berrio2006environmental}
Berr{\'\i}o JC, Hooghiemstra H, van Geel B, Ludlow-Wiechers B (2006)
  Environmental history of the dry forest biome of guerrero, mexico, and human
  impact during the last c. 2700 years. The Holocene 16(1):63

\bibitem[{Boyer(2008)}]{boyer2008intricate}
Boyer D (2008) Intricate dynamics of a deterministic walk confined in a strip.
  EPL (Europhysics Letters) 83:20,001

\bibitem[{Boyer and Larralde(2005)}]{boyer2005looking}
Boyer D, Larralde H (2005) Looking for the right thing at the right place:
  Phase transition in an agent model with heterogeneous spatial resources.
  Complexity 10(3):52--55

\bibitem[{Boyer and Walsh(2010)}]{boyer2010modelling}
Boyer D, Walsh PD (2010) Modelling the mobility of living organisms in
  heterogeneous landscapes: does memory improve foraging success? Philosophical
  Transactions of the Royal Society A: Mathematical, Physical and Engineering
  Sciences 368(1933):5645

\bibitem[{Boyer et~al(2004)Boyer, Miramontes, Ramos-Fern{\'a}ndez, Mateos, and
  Cocho}]{boyer2004modeling}
Boyer D, Miramontes O, Ramos-Fern{\'a}ndez G, Mateos JL, Cocho G (2004)
  Modeling the searching behavior of social monkeys. Physica A: Statistical
  Mechanics and its Applications 342(1-2):329--335

\bibitem[{Boyer et~al(2006)Boyer, Ramos-Fern{\'a}ndez, Miramontes, Mateos,
  Cocho, Larralde, Ramos, and Rojas}]{boyer2006scale}
Boyer D, Ramos-Fern{\'a}ndez G, Miramontes O, Mateos JL, Cocho G, Larralde H,
  Ramos H, Rojas F (2006) Scale-free foraging by primates emerges from their
  interaction with a complex environment. Proceedings of the Royal Society B:
  Biological Sciences 273(1595):1743

\bibitem[{Boyer et~al(2008)Boyer, Miramontes, and
  Ramos-Fern{\'a}ndez}]{boyer2008evidence}
Boyer D, Miramontes O, Ramos-Fern{\'a}ndez G (2008) Evidence for biological
  l\'evy flights stands. Arxiv preprint arXiv:08021762

\bibitem[{Boyer et~al(2009)Boyer, Miramontes, and Larralde}]{boyer2009levy}
Boyer D, Miramontes O, Larralde H (2009) L{\'e}vy-like behaviour in
  deterministic models of intelligent agents exploring heterogeneous
  environments. Journal of Physics A: Mathematical and Theoretical 42:434,015

\bibitem[{Boyer et~al(2011)Boyer, Miramontes, and
  Bartumeus}]{miramontes2011fractal}
Boyer D, Miramontes O, Bartumeus F (2011) The effects of spatially
  heterogeneous prey distributions on detection patterns in foraging seabirds.
  Manuscript

\bibitem[{Brantingham(2006)}]{brantingham2006measuring}
Brantingham PJ (2006) Measuring forager mobility. Current anthropology
  47(3):435--459

\bibitem[{Brockmann et~al(2006)Brockmann, Hufnagel, and
  Geisel}]{brockmann2006scaling}
Brockmann D, Hufnagel L, Geisel T (2006) The scaling laws of human travel.
  Nature 439(7075):462--465

\bibitem[{Brown et~al(2007)Brown, Liebovitch, and Glendon}]{brown2007levy}
Brown CT, Liebovitch LS, Glendon R (2007) L{\'e}vy flights in dobe ju/’hoansi
  foraging patterns. Human Ecology 35(1):129--138

\bibitem[{Brown et~al(2010)Brown, Liebovitch, and Glendon}]{brown2010hunter}
Brown CT, Liebovitch LS, Glendon R (2010) Hunter-gatherers optimize their
  foraging patterns using l\'evy flights. In: Bates DG, Tucker J (eds) Human
  Ecology, Springer US, pp 51--65

\bibitem[{Burke et~al(2009)Burke, Miguel, Satyanath, Dykema, and
  Lobell}]{burke2009warming}
Burke MB, Miguel E, Satyanath S, Dykema JA, Lobell D (2009) Warming increases
  the risk of civil war in africa. Proceedings of the National Academy of
  Sciences 106(49):20,670

\bibitem[{Burnham and Anderson(2002)}]{burnham2002model}
Burnham KP, Anderson DR (2002) Model selection and multimodel inference: a
  practical information-theoretic approach. Springer Verlag

\bibitem[{Camacho(2007)}]{camacho2007montana}
Camacho Z (2007) Monta\~{n}a de guerrero pobreza y militarizaci\'on. Revista
  Contral\'inea A\~no 5(70)

\bibitem[{Campiteli et~al(2006)Campiteli, Martinez, and
  Bruno}]{campiteli2006image}
Campiteli M, Martinez A, Bruno O (2006) An image analysis methodology based on
  deterministic tourist walks. Advances in Artificial
  Intelligence-IBERAMIA-SBIA 2006 pp 159--167

\bibitem[{Cervantes et~al(1998)Cervantes, Arriaga, Meave, and
  Carabias}]{cervantes1998growth}
Cervantes V, Arriaga V, Meave J, Carabias J (1998) Growth analysis of nine
  multipurpose woody legumes native from southern mexico. Forest Ecology and
  Management 110(1-3):329--341

\bibitem[{Charnov(1976)}]{charnov1976optimal}
Charnov EL (1976) Optimal foraging, the marginal value theorem. Theoretical
  population biology 9(2):129--136

\bibitem[{Chechkin et~al(2008)Chechkin, Metzler, Klafter, and
  Gonchar}]{chechkin2008intro}
Chechkin AV, Metzler R, Klafter J, Gonchar VY (2008) Introduction to the theory
  of l\'evy flights. In: Klages R, Radons G, Sokolov IM (eds) Anomalous
  Transport: Foundations and Applications, Wiley-VCH, Berlin


\bibitem[{Edwards et~al(2007)Edwards, Phillips, Watkins, Freeman, Murphy,
  Afanasyev, Buldyrev, da~Luz, Raposo, Stanley et~al}]{edwards2007revisiting}
Edwards AM, Phillips RA, Watkins NW, Freeman MP, Murphy EJ, Afanasyev V,
  Buldyrev SV, da~Luz MGE, Raposo EP, Stanley HE, et~al (2007) Revisiting
  l{\'e}vy flight search patterns of wandering albatrosses, bumblebees and
  deer. Nature 449(7165):1044--1048

\bibitem[{Enquist and Niklas(2001)}]{enquist2001invariant}
Enquist B, Niklas K (2001) Invariant scaling relations across tree-dominated
  communities. Nature 410(6829):655--660

\bibitem[{Flores(2007)}]{flores2007dispersal}
Flores JC (2007) Dispersal time for ancient human migrations: Americas and
  europe colonization. EPL (Europhysics Letters) 79:18,004

\bibitem[{Ghilardi et~al(2007)Ghilardi, Guerrero, and Masera}]{Ghilardi2007475}
Ghilardi A, Guerrero G, Masera M (2007) Spatial analysis of residential
  fuelwood supply and demand patterns in mexico using the wisdom approach.
  Biomass and Bioenergy 31(7):475 -- 491

\bibitem[{Gonzalez et~al(2008)Gonzalez, Hidalgo, and
  Barab{\'a}si}]{gonzalez2008understanding}
Gonzalez MC, Hidalgo CA, Barab{\'a}si AL (2008) Understanding individual human
  mobility patterns. Nature 453(7196):779--782

\bibitem[{Hays et~al(2011)Hays, Bastian, Doyle, Fossette, Gleiss, Gravenor,
  Hobson, Humphries, Lilley, Pade et~al}]{hays2011high}
Hays GC, Bastian T, Doyle TK, Fossette S, Gleiss AC, Gravenor MB, Hobson VJ,
  Humphries NE, Lilley MKS, Pade NG, et~al (2011) High activity and l{\'e}vy
  searches: jellyfish can search the water column like fish. Proceedings of the
  Royal Society B: Biological Sciences

\bibitem[{H{\'e}bert(2006)}]{hebert2006ni}
H{\'e}bert M (2006) Ni la guerre, ni la paix: campagnes de ``stabilisation" et
  violence structurelle chez les tlapan\'eques de la monta{\~n}a du guerrero
  (mexique). Anthropologica 48(1):29--42

\bibitem[{Homer-Dixon(1994)}]{homer1994environmental}
Homer-Dixon TF (1994) Environmental scarcities and violent conflict: evidence
  from cases. International security 19(1):5--40

\bibitem[{Hsiang et~al(2011)Hsiang, Meng, and Cane}]{Hsiang2011civil}
Hsiang SM, Meng KC, Cane MA (2011) Civil conflicts are associated with the
  global climate. Nature 476(49):438–441

\bibitem[{IEA(2009)}]{international2009world}
IEA (2009) International Energy Agency - 2009 World Energy Outlook. OECD/IEA,
  Paris

\bibitem[{de~Jager et~al(2011)de~Jager, Weissing, Herman, Nolet, and van~de
  Koppel}]{de2011levy}
de~Jager M, Weissing FJ, Herman PMJ, Nolet BA, van~de Koppel J (2011) L{\'e}vy
  walks evolve through interaction between movement and environmental
  complexity. Science 332(6037):1551

\bibitem[{James et~al(2011)James, Plank, and Edwards}]{james2011assessing}
James A, Plank MJ, Edwards AM (2011) Assessing l{\'e}vy walks as models of
  animal foraging. Journal of The Royal Society Interface

\bibitem[{Klafler and Sokolov(2005)}]{klafler2005anomalous}
Klafler J, Sokolov IM (2005) Anomalous diffusion spreads its wings. Physics
  world 18(8):29

\bibitem[{Knoppien and Reddingius(1985)}]{Knoppien1985273}
Knoppien P, Reddingius J (1985) Predators with two modes of searching: A
  mathematical model. Journal of Theoretical Biology 114(2):273 -- 301

\bibitem[{Landa et~al(1997)Landa, Meave, and Carabias}]{landa1997}
Landa R, Meave J, Carabias J (1997) Environmental deterioration in rural
  mexico: an examination of the concept. Ecological Applications 7(1):316--329

\bibitem[{Lima et~al(2001)Lima, Martinez, and Kinouchi}]{lima2001deterministic}
Lima GF, Martinez AS, Kinouchi O (2001) Deterministic walks in random media.
  Physical Review Letters 87(1):10,603

\bibitem[{Lomholt et~al(2008)Lomholt, Tal, Metzler, and
  Joseph}]{lomholt2008lévy}
Lomholt MA, Tal K, Metzler R, Joseph K (2008) L\'evy strategies in intermittent
  search processes are advantageous. Proceedings of the National Academy of
  Sciences 105(32):11,055

\bibitem[{Mao and Bian(2010)}]{mao2010dynamic}
Mao L, Bian L (2010) A dynamic network with individual mobility for designing
  vaccination strategies. Transactions in GIS 14(4):533--545

\bibitem[{Matth{\"a}us et~al(2011)Matth{\"a}us, Mommer, Curk, and
  Dobnikar}]{matthaus2011origin}
Matth{\"a}us F, Mommer MS, Curk T, Dobnikar J (2011) On the origin and
  characteristics of noise-induced l{\'e}vy walks of e. coli. PloS one
  6(4):e18,623


\bibitem[{Miramontes and Luque(2002)}]{miramontes2002dynamical}
Miramontes O, Luque B (2002) Dynamical small-world behavior in an epidemical
  model of mobile individuals. Physica D: Nonlinear Phenomena 168:379--385

\bibitem[{Morales-Hern\'andez(2006)}]{morales2006remesas}
Morales-Hern\'andez R (2006) Remesas familiares y condiciones de vida en el
  contexto de la migraci\' on guerrerense hacia los Estados Unidos de Am\'erica
  (PhD Thesis). Universidad Aut\'onoma de Guerrero

\bibitem[{Newman(2005)}]{newman2005power}
Newman MEJ (2005) Power laws, pareto distributions and zipf’s law.
  Contemporary Physics 46(5):323--351

\bibitem[{Nurzaman et~al(2011)Nurzaman, Matsumoto, Nakamura, Shirai, Koizumi,
  and Ishiguro}]{Nurzaman2011from}
Nurzaman SG, Matsumoto Y, Nakamura Y, Shirai K, Koizumi S, Ishiguro H (2011)
  From l\'evy to brownian: A computational model based on biological
  fluctuation. PLoS ONE 6(2):e16,168

\bibitem[{{R Development Core Team}(2009)}]{rmanual}
{R Development Core Team} (2009) R: A Language and Environment for Statistical
  Computing. R Foundation for Statistical Computing, Vienna, Austria.

\bibitem[{Raleigh and Urdal(2007)}]{raleigh2007climate}
Raleigh C, Urdal H (2007) Climate change, environmental degradation and armed
  conflict. Political Geography 26(6):674--694

\bibitem[{Ramos-Fern{\'a}ndez et~al(2004)Ramos-Fern{\'a}ndez, Mateos,
  Miramontes, Cocho, Larralde, and Ayala-Orozco}]{ramos2004lévy}
Ramos-Fern{\'a}ndez G, Mateos J, Miramontes O, Cocho G, Larralde H,
  Ayala-Orozco B (2004) L{\'e}vy walk patterns in the foraging movements of
  spider monkeys (ateles geoffroyi). Behavioral Ecology and Sociobiology
  55(3):223--230

\bibitem[{Ramos-Fern{\'a}ndez et~al(2006)Ramos-Fern{\'a}ndez, Boyer, and
  G{\'o}mez}]{ramos2006complex}
Ramos-Fern{\'a}ndez G, Boyer D, G{\'o}mez VP (2006) A complex social structure
  with fission--fusion properties can emerge from a simple foraging model.
  Behavioral ecology and sociobiology 60(4):536--549

\bibitem[{Reynolds(2008)}]{reynolds2008deterministic}
Reynolds AM (2008) Deterministic walks with inverse-square power-law scaling
  are an emergent property of predators that use chemotaxis to locate randomly
  distributed prey. Physical Review E 78(1):011,906

\bibitem[{Rhee et~al(2011)Rhee, Shin, Hong, Lee, Kim, and Chong}]{rhee2011levy}
Rhee I, Shin M, Hong S, Lee K, Kim SJ, Chong S (2011) On the levy-walk nature
  of human mobility. IEEE-ACM Transactions on Networking 19(3):630--643

\bibitem[{Richerson and Boyd(2008)}]{richerson2008being}
Richerson PJ, Boyd R (2008) Being human: Migration: An engine for social
  change. Nature 456(7224):877--877

\bibitem[{Santos et~al(2007)Santos, Boyer, Miramontes, Viswanathan, Raposo,
  Mateos, and Da~Luz}]{santos2007origin}
Santos MC, Boyer D, Miramontes O, Viswanathan GM, Raposo EP, Mateos JL, Da~Luz
  MGE (2007) Origin of power-law distributions in deterministic walks: The
  influence of landscape geometry. Physical Review E 75(6):061,114

\bibitem[{Schreier and Grove(2010)}]{Schreier201075}
Schreier AM, Grove M (2010) Ranging patterns of hamadryas baboons: random walk
  analyses. Animal Behaviour 80(1):75 -- 87

\bibitem[{Shlesinger and Klafter(2000)}]{shlesinger1986lévy}
Shlesinger MF, Klafter J (2000) L\'evy walks versus l\'evy flights. In: Stanley
  HE, Ostrowsky N (eds) On Growth and Form: Fractal and Non-Fractal Patters in
  Physics, Martinus Nijhoff Publishers, Dordrecht, pp 279--283

\bibitem[{Sims et~al(2008)Sims, Southall, Humphries, Hays, Bradshaw, Pitchford,
  James, Ahmed, Brierley, Hindell et~al}]{sims2008scaling}
Sims DW, Southall EJ, Humphries NE, Hays GC, Bradshaw CJA, Pitchford JW, James
  A, Ahmed MZ, Brierley AS, Hindell MA, et~al (2008) Scaling laws of marine
  predator search behaviour. Nature 451(7182):1098--1102

\bibitem[{Slatkin(1973)}]{slatkin1973gene}
Slatkin M (1973) Gene flow and selection in a cline. Genetics 75(4):733

\bibitem[{Sokal et~al(1989)Sokal, Harding, and Oden}]{sokal1989spatial}
Sokal RR, Harding RM, Oden NL (1989) Spatial patterns of human gene frequencies
  in europe. American journal of physical anthropology 80(3):267--294

\bibitem[{Stanley and Buldyrev(2001)}]{stanley2001salesman}
Stanley HE, Buldyrev SV (2001) The salesman and the tourist. Nature 413:373

\bibitem[{Stephens and Krebs(1986)}]{stephens1986foraging}
Stephens DW, Krebs JR (1986) Foraging theory. Princeton Univ Pr

\bibitem[{Sueur(2011)}]{sueur2011non}
Sueur C (2011) A non-l{\'e}vy random walk in chacma baboons: What does it mean?
  PloS one 6(1):e16,131

\bibitem[{Taniguchi(2011)}]{taniguchi2011guerrero}
Taniguchi H (2011) Guerrero tiene municipios tan pobres como algunos pa\'ises
  de africa. CNN M\'exico. http://mexico.cnn.com/

\bibitem[{Viswanathan et~al(1999)Viswanathan, Buldyrev, Havlin, Da~Luz, Raposo,
  and Stanley}]{viswanathan1999optimizing}
Viswanathan GM, Buldyrev SV, Havlin S, Da~Luz MGE, Raposo EP, Stanley HE (1999)
  Optimizing the success of random searches. Nature 401(6756):911--914

\bibitem[{Viswanathan et~al(2011)Viswanathan, da~Luz, Raposo, and
  Stanley}]{viswanathan2011physics}
Viswanathan GM, da~Luz MGE, Raposo EP, Stanley HE (2011) The Physics of
  Foraging: An Introduction to Random Searches and Biological Encounters.
  Cambridge Univ Pr

\bibitem[{Wakeley(1999)}]{wakeley1999nonequilibrium}
Wakeley J (1999) Nonequilibrium migration in human history. Genetics
  153(4):1863


%
%

\end{thebibliography}


\end{document}